\documentclass[prd,a4paper,showpacs,preprintnumbers,nofootinbib]{revtex4} 

\usepackage{epsfig}

\def\nn{\nonumber}        

\newcommand{\bm}[1]{\mbox{\boldmath $#1$}}

\newcommand{\open}{{<\kern -0.3 em{\scriptscriptstyle )}}}

\newcommand{\nslash}{\kern 0.2 em n\kern -0.45em /}
\newcommand{\Pslash}{\kern 0.2 em P\kern -0.56em \raisebox{0.3ex}{/}}
\newcommand{\pslash}{\kern 0.2 em p\kern -0.4em /}
\newcommand{\kslash}{\kern 0.2 em k\kern -0.45em /}
\newcommand{\Sslash}{\kern 0.2 em S\kern -0.56em \raisebox{0.3ex}{/}}

\newcommand{\eq}{\begin{equation}}
\newcommand{\ee}{\end{equation}}
\newcommand{\beq}{\begin{equation}}
\newcommand{\eeq}{\end{equation}}
\newcommand{\ba}{\begin{eqnarray}}
\newcommand{\ea}{\end{eqnarray}}
\newcommand{\eqa}{\begin{eqnarray}}
\newcommand{\eea}{\end{eqnarray}}

\newcommand{\sumint}{\kern 0.2 em {\textstyle\sum} \kern -1.1 em \int}

\newcommand{\la}{\langle}
\newcommand{\ra}{\rangle}
\newcommand{\amp}[1]{\la #1 \ra}
\newcommand{\fapi}{\frac{\alpha_s}{2\pi}}

\begin{document} 

\title{Drell-Yan Lepton Angular Distribution at Small Transverse Momentum}

\author{Dani\"el Boer}
\email{D.Boer@few.vu.nl}
\affiliation{Dept.\ of Physics and Astronomy, Vrije Universiteit Amsterdam, \\
De Boelelaan 1081, 1081 HV Amsterdam, The Netherlands}

\author{Werner Vogelsang}
\email{vogelsan@quark.phy.bnl.gov}
\affiliation{Physics Department, 
Brookhaven National Laboratory, Upton, NY 11973, U.S.A.}

\date{\today}

\begin{abstract}
We investigate the dependence of the Drell-Yan cross section
on lepton polar and azimuthal angles, as generated by the lowest-order
QCD annihilation and Compton processes. We focus in particular
on the azimuthal-angular distributions, which are
of the form $\cos\phi$ and $\cos 2\phi$. At small transverse momentum $q_T$ 
of the lepton pair, $q_T\ll Q$, with $Q$ the pair mass, these terms are known 
to be suppressed relative to the $\phi$-independent part of the Drell-Yan
cross section by one or two powers of the transverse momentum. Nonetheless, as
we show, like the $\phi$-independent part they are subject to large 
logarithmic corrections, whose precise form however depends on the reference 
frame chosen. These logarithmic contributions ultimately require resummation
to all orders in the strong coupling. We discuss the potential effects of 
resummation on the various angular terms in the cross section and on 
the Lam-Tung relation. 
\end{abstract}

\preprint{BNL-NT-06/16} 


\maketitle


\section{Introduction}
It is well known that the angular distribution of the leptons 
in the Drell-Yan process $P_1 \, P_2 \to \ell \, \bar{\ell} \, X$ may possess
azimuthal asymmetries. These correspond to angular-dependent terms in the 
ratio of differential cross sections,
\beq
\frac{dN}{d\Omega} \equiv \left(\frac{d \sigma}{d^4 q}
\right)^{-1} \frac{d \sigma}{d\Omega d^4 q} \; , \label{angdef}
\eeq
where $q$ is the four-momentum of the virtual photon (or $Z$ boson, if
the energy is sufficiently high) decaying into the lepton pair, and 
$d\Omega\equiv d\cos\theta d\phi$ is the solid angle of the lepton $\ell$
in terms of its polar and azimuthal angles in the center-of-mass system
(c.m.s.) of the lepton pair. As we shall review below, an analysis 
of the general Lorentz structure of the hadronic tensor yields
the following angular structure~\cite{oakes,Lam-78}:
\ba
\frac{dN}{d\Omega}  =  
\frac{3}{8\pi} \; \frac{W_T (1+ \cos^2 \theta)   + W_L (1- \cos^2\theta) 
+ W_\Delta \sin 2\theta \cos \phi + W_{\Delta \Delta} \sin^2 \theta 
\cos 2\phi}{2 W_T + W_L} \; . \label{sigmaW}
\ea
Here the ``structure functions'' $W_{T,L,\Delta,\Delta \Delta}$ 
depend on the virtual photon's invariant mass $Q$, its transverse momentum 
$Q_T$, and its rapidity $y$. Within the lepton pair c.m.s.\ there is still
freedom to choose the axes of the coordinate system, with respect to which
the lepton angles are defined. The $W_{T,L,\Delta,\Delta \Delta}$ also depend
on this choice. Equivalently to Eq.~(\ref{sigmaW}), one also 
often writes the lepton angular distribution as 
\ba
\frac{dN}{d\Omega} & = & 
\frac{3}{4\pi} \; \frac{1}{\lambda+3} \left[ 1+ \lambda \cos^2\theta
+ \mu \sin 2\theta \cos \phi + \frac{\nu}{2} \sin^2 \theta 
\cos 2\phi \right].
\label{DNdOmega}
\ea
One obviously has
\ba
\lambda=\frac{W_T-W_L}{W_T+W_L}\; , \;\;\;
\mu=\frac{W_\Delta}{W_T+W_L}\; , \;\;\;
\nu=\frac{2W_{\Delta\Delta}}{W_T+W_L} \; .
\label{lmnWrel}
\ea
A third, again equivalent, parameterization for the lepton angular
distribution~\cite{CS-77} that is also often employed is 
\ba
\frac{dN}{d\Omega} & = & 
\frac{3}{16\pi} \left[ 1+ \cos^2\theta + \frac{A_0}{2} (1-3
\cos^2\theta) 
+ A_1 \sin 2\theta \cos \phi + \frac{A_2}{2} \sin^2 \theta \cos 2\phi \right] \; .
\label{Adef}
\ea
Evidently, 
\beq
\lambda = \frac{2-3 A_0}{2+A_0} \ , \quad
\mu = \frac{2A_1}{2+A_0} \ , \quad
\nu = \frac{2A_2}{2+A_0} \ ,
\eeq
or  
\beq
A_0 = \frac{2W_L}{2W_T+W_L} \ , \quad 
A_1 = \frac{2W_{\Delta}}{2W_T+W_L} \ , \quad 
A_2 = \frac{4W_{\Delta \Delta}}{2W_T+W_L} \ .
\eeq

If $Q$ is large, one may calculate the structure functions using parton
model concepts, factorizing them into collinear convolutions of the parton distributions 
of the two scattering hadrons and partonic hard-scattering cross sections that 
are amenable to QCD perturbation theory. For non-vanishing transverse momentum
$Q_T$ of the virtual photon, the lowest-order (LO) partonic processes are
$q\bar{q}\to \gamma^* g$ and $qg\to \gamma^* q$. The contributions to
the angular distributions by these processes have been calculated in 
Refs.~\cite{Kajantie-78,Lam-79,Collins-79,Cleymans-78,Lindfors-79}. Also
the next-to-leading order (NLO) corrections have been 
derived~\cite{Mirkes:1992hu,Mirkes:1994dp}. 

Experimentally, for a given lepton pair invariant mass $Q$, the bulk of 
the Drell-Yan events is at rather low transverse momenta of the pair, 
$Q_T\ll Q$. It is this regime that is also most interesting from a 
theoretical point of view. ``Intrinsic'' transverse momenta of the
initial partons may become relevant. Also, as $Q_T\ll Q$, fixed-order 
calculations of the partonic cross sections are bound to fail. When $Q_T\to 0$, 
gluon radiation is inhibited, so that only relatively soft gluons may be
emitted into the final state. The cancellation of infra-red singularities 
between real and virtual diagrams in the perturbative series then leaves 
behind large logarithmic remainders of the form
$\alpha_s^k\,\ln^{m}\left(Q^2 /Q_T^2\right)/Q_T^2$
in the cross section $d\sigma/d^4q$
at the $k$th order of perturbation theory, where $m=1,\ldots, 2k-1$. Ultimately,
when $Q_T\ll Q$, $\alpha_s$ will not be useful anymore as the expansion 
parameter in the perturbative series since the logarithms will compensate
for the smallness of $\alpha_s$. Accordingly, in order to obtain a 
reliable estimate for the cross section, one has to sum up (``resum'') 
the large logarithmic contributions to all orders in $\alpha_s$.
Techniques for this resummation are well established, starting
with pioneering work mostly on the Drell-Yan process in the late 
1970's to mid 1980's~\cite{DDT78,PP79,Ellis-80,Ellis-81,CS81,CS82,AEGM84,CSS85}. 
The ``Collins-Soper-Sterman'' (CSS) formalism~\cite{CSS85} has become the 
standard method for $Q_T$ resummation. 
It is formulated in 
impact-parameter ($b$) space, which guarantees conservation of the 
transverse momenta of the emitted soft gluons.

The large terms $\alpha_s^k\,\ln^{m}\left(Q^2 /Q_T^2\right)/Q_T^2$
just described occur only in the part proportional to $W_T$ of the cross
section $d\sigma/d\Omega d^4 q$. The function $W_\Delta$ is less singular by one
power of $Q_T$ at small $Q_T$, while $W_L$ and $W_{\Delta\Delta}$ are
suppressed even by a factor $Q_T^2$ relative to $W_T$. For $W_\Delta$ and
$W_{\Delta\Delta}$ this can be
understood as follows: at $Q_T = 0$ the angle $\phi$ cannot be
defined, hence no azimuthal asymmetry could be observable, and therefore, the
asymmetries should smoothly go to zero in the limit of $Q_T \to 0$.

If one is just interested in the small-$Q_T$ behavior of the cross section 
$d\sigma/d^4 q$, it will be sufficient to take into account the resummation 
of the large logarithms in $W_T$. However, this may become different if one
considers the parameters $\lambda,\mu,\nu$ or $A_{0,1,2}$ defined
above, which were the object of dedicated experimental studies in 
$\pi^- N$~Drell-Yan experiments about 
20 years ago~\cite{Falciano,Guanziroli,Conway} (experimental evidence for
nonzero $W_L$ was already reported in~\cite{Anderson-79}).
Superficially, one might think that, in order to improve the 
theoretical prediction for these coefficients by resummation, it will be 
sufficient to perform the resummation in $W_T$ alone. This is in fact what 
has been done in the literature so far in~\cite{Chiappetta-86} for
Drell-Yan and in~\cite{Meng:1995yn,Nadolsky-99,Nadolsky-00}
for the related ``semi-inclusive deeply-inelastic scattering'' (SIDIS)
process and has led, for example, to the claim in Ref.~\cite{Chiappetta-86} 
that resummation has a strong effect on the perturbative-QCD results for
$\lambda, \mu$ and $\nu$ at small $Q_T$. 
However, even though $W_L$, $W_\Delta$, and $W_{\Delta\Delta}$ are
down by powers of $Q_T$ as $Q_T\to 0$, they all individually may receive very
similar large logarithmic corrections at small $Q_T$ as $W_T$ does.
Therefore, it may well happen that in a ratio such as $A_2=4 W_{\Delta
  \Delta}/(2 W_T+W_L)$ 
the resummation effects cancel to a large degree, if not
completely. In the present paper, we will investigate the small-$Q_T$ 
behavior of $W_L$, $W_\Delta$, and $W_{\Delta\Delta}$, and we will also 
address some qualitative consequences for the $Q_T$-resummation for these 
and for the various coefficients constructed from them. A complication arises
from the fact that the $W_L$, $W_\Delta$, and $W_{\Delta\Delta}$
depend on the coordinate frame chosen. In a change of frame, 
terms proportional to $Q_T$ or $Q_T^2$ may be redistributed 
among $W_T$ and the $W_{L,\Delta,\Delta\Delta}$ and, 
because $W_T$ is more singular at small $Q_T$, 
may alter the small-$Q_T$ behavior of $W_{L,\Delta,\Delta\Delta}$. Our
aim is to exhibit this feature very explicitly by considering two
particular frames commonly used in the literature, in order to eliminate
some misconceptions concerning the small-$Q_T$ behavior and the 
resummation of angular distributions. 

As discussed above, the CSS formalism~\cite{CSS85} describes the behavior of 
the structure function $W_T$ at small $Q_T$. It may be used 
to predict the $Q_T$-singular pieces arising at a given perturbative order.
To LO, these have a particularly simple structure~\cite{AEGM84}, involving 
the first-order expansion of the Sudakov form factor and the LO 
DGLAP~\cite{ap} splitting functions. This structure is straightforwardly 
recovered from explicit calculations of the cross sections for the 
partonic reactions $q\bar{q}\to \gamma^* g$ and $qg\to \gamma^* q$.
When we use these processes to determine the small-$Q_T$ behavior
of the other structure functions $W_L$, $W_\Delta$, and $W_{\Delta\Delta}$,
we find a closely related, but different, form that involves 
different ``splitting functions''. This is likely due to the fact that
the {\it direction} of the observed transverse momentum matters for these
structure functions. It is important to emphasize that the CSS formalism 
was not constructed to treat the directional dependence
of the transverse momentum distribution. Even though we will discuss in this 
paper
some features of the resummation of the  $W_{L,\Delta,\Delta\Delta}$, 
we have not been able to organize their small-$Q_T$ behavior beyond
the leading logarithms into a resummed form in terms of an ``extended'' 
CSS formalism that goes beyond collinear factorization. 
We therefore hope that the study presented in this paper 
will serve as a motivation for a more general analysis of resummation in 
cases where the direction of the observed transverse momentum matters. In 
this respect 
the proper starting point will be to consider the factorizations and
techniques put forward in Refs.\ \cite{CS81,CS82,JiMaYuan}.      

The article is organized as follows. In Section II we express the
azimuthal asymmetries in terms of structure functions in two different frames
that are commonly used in the literature. This analysis is exact in $Q_T$ and
does not rely on the use of perturbative QCD, just on kinematical considerations. 
In Section III we discuss the LO perturbative QCD results, and in Section IV we
investigate their small-$Q_T$ limit. In Section V we address
some aspects of the resummation of the angular terms in this limit. 
We present our conclusions in Section VI.  

\section{Azimuthal dependences in terms of structure functions}
In this section we discuss in detail the $\phi$ dependence of the differential 
cross section in terms of the structure functions $W_{T,L,\Delta,\Delta\Delta}$. 
This is a purely kinematical analysis, which will set the notation to be used 
later on. We follow (in part) the notation of Refs.\ \cite{Lam-78,AL-82}, 
but use the metric ${\rm diag} (+ - - \; -)$. In passing we will
point out some differences with results that appeared in the earlier 
literature. 
 
We first define invariant 
structure functions $W_{1,2,3,4}$ through the following 
parameterization of the hadronic tensor:
\beq
{W}^{\mu \nu} = - (g^{\mu \nu}- \frac{q^\mu q^\nu}{q^2}) W_1 +
\tilde{P}^\mu \tilde{P}^\nu W_2 - \frac{1}{2} (\tilde{P}^\mu \tilde{p}^\nu + 
\tilde{p}^\mu \tilde{P}^\nu)W_3 + \tilde{p}^\mu \tilde{p}^\nu W_4 \; ,
\label{W1234}
\eeq
where $P=P_1+P_2$, $p=P_1-P_2$, $\tilde{P}^\mu = (P^\mu - q^\mu (P\cdot
q)/Q^2)/\sqrt{s}$, $\tilde{p}^\mu = (p^\mu - q^\mu (p\cdot q)/Q^2)/\sqrt{s}$,
with $P_{1,2}$ the initial hadron momenta and $q$ the momentum of the virtual photon,
$Q^2\equiv q^2$. The hadronic c.m.s.\ energy is $\sqrt{s}=\sqrt{(P_1+P_2)^2}$.

As mentioned in the Introduction, we will use two different reference 
frames: the so-called Collins-Soper (CS) frame~\cite{CS-77} and 
the Gottfried-Jackson (GJ) frame~\cite{Lam-78} (also sometimes 
referred to as the ``$t$-channel helicity frame''). Both frames are
rest frames of the virtual photon or, equivalently, of the lepton pair. 
However, this property does not completely 
specify a frame, as one has a freedom in the choice of the coordinate axes,
corresponding to rotations among the frames. In both the CS and GJ
frames, a set of orthonormal axes ($X$, $Y$, $Z$ and $T$) are defined,
and the hadronic tensor is re-expressed as:
\beq
{W}^{\mu \nu} = - (g^{\mu \nu}- T^\mu T^\nu) (W_T+ W_{\Delta
\Delta}) - 2 X^\mu X^\nu W_{\Delta \Delta} + Z^\mu
Z^\nu (W_L -W_T - W_{\Delta \Delta} ) - (X^\mu Z^\nu +
Z^\mu X^\nu) W_\Delta \ .
\label{WTLDDD}
\eeq
In this way one has $W^\mu{}_\mu = -(2W_T + W_L)$, which is the quantity
that appears in the differential cross section $d\sigma/d^4 q$, the denominator of
$dN/d\Omega$ in Eq.~(\ref{angdef}). Contracting with the straightforwardly 
calculated leptonic tensor, one finds the cross section in terms of the 
structure functions: 
\beq
\frac{d\sigma}{d\Omega d^4 q}=
\frac{\alpha^2}{2(2\pi)^4Q^2s^2}\;\left\{ 
W_T(1+ \cos^2 \theta) + W_L (1- \cos^2\theta)
+ W_\Delta \sin 2\theta \cos \phi + W_{\Delta \Delta} \sin^2 \theta
\cos 2\phi \right\} \ , \label{crsec}
\eeq
again valid in both frames. Here $\alpha$ is the electromagnetic
coupling constant. From Eq.~(\ref{crsec}) one immediately reproduces 
Eq.~(\ref{sigmaW}). Higher harmonics in $\cos (n\phi)$ do not occur in the 
angular distribution due to the fact that the kinematics 
of the process are fully determined by only three momentum vectors, 
$P_1$, $P_2$, and $q$. In case the polarization of the leptons is
not summed over or in case of electroweak corrections
$\sin \phi$ and $\sin 2\phi$ terms will also be present, but these will not be 
considered here. 
We also note that the structure functions $W_{T,L,\Delta,\Delta\Delta}$
are associated with specific polarizations of the 
virtual photon~\cite{Lam-78}:
$W_T=W^{1,1}$, $W_L=W^{0,0}$, $W_\Delta=(W^{0,1}+W^{1,0})/\sqrt{2}$, 
and $W_{\Delta \Delta}=W^{1,-1}$, where the first (second) superscript denotes
the photon helicity in the amplitude (its complex conjugate) of the 
Drell-Yan process. The azimuthal dependence introduced by $W_\Delta$ and
$W_{\Delta \Delta}$ therefore comes from their single- or double-spin-flip 
property, respectively. 

\begin{figure}[b]
\begin{center}
\leavevmode \epsfxsize=9cm \epsfbox{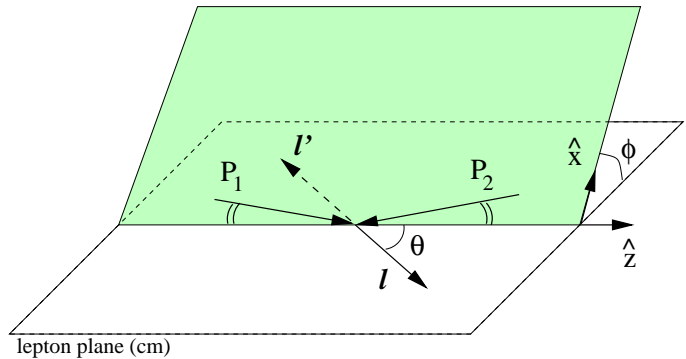}
\vspace{0.2 cm}
\caption{\label{CSframe} The Collins-Soper frame.}
\end{center}
\end{figure}

In both frames, $T^\mu \equiv q^\mu/Q$ and $Y^\mu \equiv 
\epsilon^{\mu\nu\alpha\beta} X_\nu Z_\alpha T_\beta$. 
In the CS frame, the $Z$ axis is defined as pointing in the
direction that bisects the angle between the 
three-vectors $\vec{P}_2$ and $-\vec{P}_1$,
see Fig.~\ref{CSframe}. This gives
\ba
Z^\mu & \equiv & \frac{1}{\sqrt{Q^2+Q_T^2}}\left(q_p \tilde{P}^\mu + q_P
  \tilde{p}^\mu \right) = \frac{2}{s \sqrt{Q^2+Q_T^2}} \left( 
(P_2\cdot q) \tilde{P}_1^\mu - (P_1\cdot q) \tilde{P}_2^\mu \right) ,
\label{CSZ}
\ea
where $Q_T^2 \equiv \bm{q}_T^2$ is the square of the transverse momentum
$\bm{q}_T$ of the virtual photon with respect to the two hadron momenta.
Furthermore, $q_P\equiv (P\cdot q)/\sqrt{s}$, $q_p\equiv -(p\cdot q)/\sqrt{s}$ 
(in the hadronic c.m.s.\ $q_P$ is the energy of the virtual photon, 
while $q_p$ is its $z$-component). Finally, in 
Eq.~(\ref{CSZ}) $\tilde{P}_i^\mu \equiv P_i^\mu - q^\mu 
(P_1\cdot q)/Q^2$ (note there is no factor $1/\sqrt{s}$ in the latter definition, 
compared to the definitions of $\tilde{P}$ and $\tilde{p}$ above). 
For the $X$-axis one chooses
\ba
X^\mu & = & -\frac{Q}{Q_T \sqrt{Q^2+Q_T^2}}\left(q_P \tilde{P}^\mu + q_p
  \tilde{p}^\mu \right) = - \frac{2Q}{s Q_T \sqrt{Q^2+Q_T^2}} \left(
(P_2\cdot q) \tilde{P}_1^\mu + (P_1\cdot q) \tilde{P}_2^\mu \right).
\label{CSX}
\ea 

\begin{figure}[h]
\begin{center}
\leavevmode \epsfxsize=9cm \epsfbox{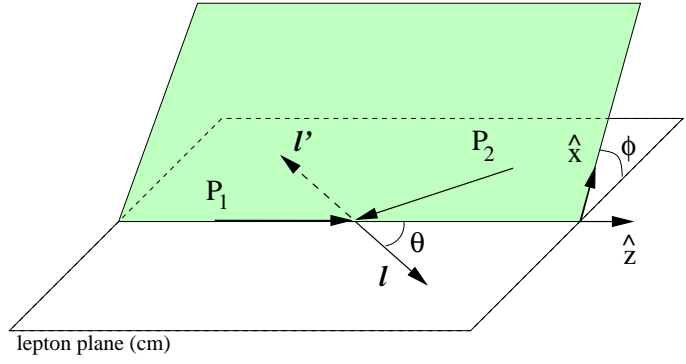}
\vspace{0.2 cm}
\caption{\label{GJframe} The Gottfried-Jackson frame.}
\end{center}
\end{figure}

In the GJ frame, the $Z$ axis points in the
direction of the three-vector $\vec{P}_1$, see Fig.~\ref{GJframe}:
\ba
Z^\mu & \equiv & \frac{Q}{q_P-q_p}\left(\tilde{P}^\mu + \tilde{p}^\mu
\right)= \frac{Q}{P_1\cdot q} \tilde{P}_1^\mu \ .
\label{GJZ}
\ea 
Furthermore,
\ba
X^\mu & = & -\frac{1}{Q_T (Q^2+Q_T^2)}\left((Q^2 q_P - Q_T^2 q_p)
  \tilde{P}^\mu + (Q^2 q_p - Q_T^2 q_P) \tilde{p}^\mu \right) =
 -\frac{2Q}{s Q_T} \left((P_2\cdot Z) 
\tilde{P}_1^\mu - (P_1\cdot Z) \tilde{P}_2^\mu \right) \ .
\label{GJX}
\ea 

One finds that the three-vector components of the $X$ and $Z$ axes of 
the two frames are related by a rotation: 
\ba
\vec{Z}_{GJ} = \cos \gamma \; \vec{Z}_{CS} + \sin \gamma \; 
\vec{X}_{CS} \ ,\\
\vec{X}_{GJ} = - \sin \gamma \; \vec{Z}_{CS} + \cos \gamma \; 
\vec{X}_{CS} \ ,
\ea
where 
\beq
\cos \gamma  = \frac{Q}{\sqrt{Q^2+Q_T^2}} \ , \quad \sin \gamma  = -
\frac{Q_T}{\sqrt{Q^2+Q_T^2}} \ .
\eeq
Thus, in the limit $Q_T \to 0$ the two sets of $X,Z$ axes coincide.

Inserting the above definitions of the coordinate axes into the hadronic 
tensor in Eq.~(\ref{WTLDDD}) and comparing to Eq.~(\ref{W1234}), one
can derive for each 
frame the relations between the sets $W_{1,2,3,4}$ and 
$W_{T,L,\Delta,\Delta\Delta}$ of structure functions. For the CS frame this 
gives (this is Eq.\ (B2) of Ref.~\cite{Lam-78}):
\ba
W_T & = & W_1 - W_{\Delta \Delta} \ , \nn \\
W_L & = & W_1 + \frac{1}{Q^2 + Q_T^2} \left( q_p^2 W_2 + q_p q_P
  W_3 + q_P^2 W_4 \right) \ ,\nn \\
W_\Delta & = & -\frac{Q_T}{Q (Q^2 + Q_T^2)} \left( q_p q_P (W_2 +
  W_4) + \frac{1}{2} (q_P^2 + q_p^2)W_3 \right) \ , \nn \\
W_{\Delta \Delta} & = & - \frac{Q_T^2}{2Q^2(Q^2 + Q_T^2)}
  \left( q_P^2 W_2 + q_p q_P W_3 + q_p^2 W_4 \right) \ ,
\label{fromitoCS}
\ea
while for the GJ frame one finds 
(this is a corrected version of Eq.\ (B3) 
of~\cite{Lam-78} and also of a corresponding expression in reference-note~5 
of~\cite{AL-82}):
\ba
W_T & = & W_1 - W_{\Delta \Delta} \ , \nn\\
W_L & = & W_1 + \left( \alpha^2 W_2 + \alpha \beta W_3 + \beta^2 W_4
  \right) \ , \nn\\
W_\Delta & = & \frac{Q_T}{q_P-q_p} \left(\alpha (W_2 + \frac{1}{2}W_3)
  + \beta (W_4 + \frac{1}{2}W_3) \right) \ , \nn\\
W_{\Delta \Delta} & = & - \frac{Q_T^2}{2(q_P - q_p)^2}
  \left( W_2 + W_3 + W_4 \right) \ ,
\label{fromitoGJ}
\ea
where 
\ba
\alpha & \equiv & \frac{q_P}{Q} - \frac{Q}{q_P - q_p} \ , \nn\\
\beta & \equiv & \frac{q_p}{Q} - \frac{Q}{q_P - q_p} \ .
\ea
The relation between the structure functions in the CS frame and the 
GJ one is given by the following linear transformation:
\beq
\left( \begin{array}{c} W_T \\ W_L \\ W_\Delta \\ W_{\Delta \Delta}
  \end{array} \right)_{GJ}
= \frac{1}{1+\rho^2} \left( \begin{array}{cccc}
1+\frac{1}{2}\rho^2 & \frac{1}{2} \rho^2 & - \rho & \frac{1}{2}\rho^2 \\
\rho^2 & 1 & 2\rho & -\rho^2 \\
\rho & - \rho & 1- \rho^2 & -\rho \\
\frac{1}{2} \rho^2 & - \frac{1}{2} \rho^2 & \rho & 1+ \frac{1}{2}\rho^2 
\end{array} \right) \; 
\left( \begin{array}{c}  W_T \\ W_L \\ W_\Delta \\ W_{\Delta \Delta}
\end{array} \right)_{CS} \ ,
\label{fromCStoGJ}
\eeq
where $\rho = Q_T/Q$. Expressed in terms of the angle $\gamma$ defined
above, the matrix in the above equation is identical to the one presented 
in~\cite{AL-82} (their Eq.\ (4)), apart from a sign in the third 
entry of the first row. 

For future reference, we also give the transformation between the 
sets of coefficients $\lambda,\mu,\nu$ in the two frames, which were defined in 
Eq.~(\ref{lmnWrel}). From Eq.\ (\ref{fromCStoGJ}) one finds:
\beq
\left( \begin{array}{c} \lambda \\ \mu \\ \nu \end{array} \right)_{GJ}
= \frac{1}{\Delta_{CS}} \left( \begin{array}{ccc}
1-\frac{1}{2}\rho^2 & - 3 \rho & \frac{3}{4}\rho^2 \\
\rho & 1- \rho^2 & -\frac{1}{2} \rho \\
\rho^2 & 2 \rho & 1+\frac{1}{2} \rho^2 
\end{array} \right) \; 
\left( \begin{array}{c} \lambda \\ \mu \\ \nu \end{array} \right)_{CS} \ ,
\label{lmnMatrix}
\eeq
where  
\beq
\Delta = 1+ \rho^2 + \frac{1}{2} \rho^2 \lambda 
+ \rho \mu - \frac{1}{4} \rho^2 \nu \ .
\eeq
The reverse transformation from the GJ frame to the CS frame is the same upon
replacement of $\rho \to - \rho$ (and exchange of the labels $CS$ and
$GJ$). This is in agreement with Ref.\ \cite{Falciano}
(note that what is referred to as ``$u$-channel'' (UC) frame in that reference
corresponds to the GJ frame used here, which accounts for the $\rho \to -
\rho$ difference with respect to our Eq.\ (\ref{lmnMatrix})).
One notes that if $\rho \to 0$, the rotation matrix becomes the unit matrix, as 
expected. All results presented so far are exact in $\rho$. 

\section{Azimuthal asymmetries from LO perturbative QCD processes}

The structure functions can be calculated employing collinear factorization 
and QCD perturbation theory. One writes down an expression analogous 
to~(\ref{W1234}) or~(\ref{WTLDDD}) for the {\it partonic} tensor, in terms of 
partonic structure functions $\widehat{W}_i$. The hadronic structure functions are
obtained as convolutions of the partonic ones with the appropriate parton
distribution functions. The $\widehat{W}_i$ may be evaluated directly from partonic 
hard-scattering processes, which at LO are the annihilation (or, 
gluon bremsstrahlung) reaction $q\bar{q}\to \gamma^* g$ and the (QCD) 
Compton process $qg\to \gamma^* q$. 
The annihilation process is the dominant process in $\pi^- p$ Drell-Yan
scattering (likewise in the $\bar{p} p$ Drell-Yan process). 
The Compton process is dominant in the Drell-Yan process in $pp$ scattering 
at large $Q_T$.

In the following, we will first focus on the annihilation process, which for
our purposes is most relevant, since the Compton process is sub-leading in the 
region of $Q_T\ll Q$ we are mostly interested in. For the process
$q\bar{q}\to \gamma^* g$ one has:
\beq 
\frac{W_i}{x_1 x_2} = \int d\xi_1\int d\xi_2 \; 
\delta\left( (\xi_1-x_1)(\xi_2-x_2) - Q_T^2/s\right) 
\sum_a e_a^2 q_a(\xi_1,\mu) \bar{q}_{a}(\xi_2,\mu) \frac{\widehat{W}_i}{\xi_1\xi_2} \ , 
\label{Wwdef}
\eeq 
where we define $x_1$ and $x_2$ by writing $q= x_1 P_1^+ +
x_2 P_2^- + q_T$, with the light-cone components of any four-vector $v$ 
given by $v^{\pm}\equiv (v^0\pm v^3)/\sqrt{2}$. We  have 
$x_1=(q_P+q_p)/\sqrt{s}$, $x_2 =(q_P-q_p)/\sqrt{s}$ with $q_P, q_p$ as
introduced in the previous section (note that in the hadronic 
c.m.s.\ one has $q_P=(s+Q^2)/2\sqrt{s}$, $q_p =
\sqrt{q_P^2-Q^2-Q_T^2}=\sqrt{(s-Q^2)^2/4s - Q_T^2}\,$).
For the partonic collinear momenta we define 
$p_1^+ = \xi_1 P_1^+$ and $p_2^- = \xi_2 P_2^-$.
The delta-function in~(\ref{Wwdef})  expresses the on-mass-shell condition
for the outgoing ``unobserved'' gluon in the process $q\bar{q}
\to \gamma^*g$.   
Equation~(\ref{Wwdef}) contains the appropriate sum over all quark and
anti-quark flavors $a$, each with their corresponding parton distributions
$q_a(x,\mu)$ and $q_{\bar{a}}(x,\mu) = \bar{q}_a(x,\mu)$, respectively, 
where $\mu\sim Q$ is the
factorization scale; $e_a^2$ is the quark's squared electromagnetic 
charge. Note that since we are only interested in the lepton 
angular distribution $dN/d\Omega$, which is a ratio of cross sections, we are
free to adjust the overall normalization of the tensors, which we do in 
such a way as to simplify the formulas. 
The annihilation process then has the following partonic structure 
functions \cite{Lam-79,Collins-79}:
\ba 
\widehat{W}_1 & = & \fapi \frac{C_F\,s}{\xi_1 \xi_2 Q_T^2} \left[\frac{Q^4}{s^2}
- \frac{2Q_T^2}{s}\xi_1\xi_2 + \xi_1^2 \xi_2^2\right] =\fapi 
\frac{C_F}{\hat{t}\hat{u}} \left[(\hat{t}-Q^2)^2+(\hat{u}-Q^2)^2 \right] \ ,\nn\\
\widehat{W}_2 & = & - \fapi \frac{C_F\,Q^2}{\xi_1 \xi_2 Q_T^2}
\left[\xi_1^2+\xi_2^2\right] = - \fapi \frac{C_F\,s Q^2}{\hat{t}\hat{u}}
\left[\xi_1^2+\xi_2^2\right] \ , \nn\\
\widehat{W}_3 & = & \frac{C_F\alpha_s}{\pi} \frac{Q^2}{\xi_1 \xi_2 Q_T^2}
\left[\xi_1^2-\xi_2^2\right] = \fapi \frac{2 C_F\,s Q^2}{\hat{t}\hat{u}}
\left[\xi_1^2-\xi_2^2\right]  \ , \nn\\
\widehat{W}_4 & = & \widehat{W}_2 \ ,
\label{Whats}
\ea
where $C_F=4/3$, and where $\hat{s}= (\xi_1 P_1 + \xi_2 P_2)^2=
\xi_1\xi_2 s$, 
$\hat{t} = (q - \xi_1 P_1)^2$ and $\hat{u} = (q-\xi_2 P_2)^2$. 
Constructing via
Eq.~(\ref{fromitoCS}) the structure functions $W_{T,L,\Delta,\Delta\Delta}$  
and inserting these into Eq.~(\ref{sigmaW}), one 
finds~\cite{Collins-79,Cleymans-78,Lindfors-79} {\em in the CS frame}:
\ba
\lefteqn{\frac{dN}{d\Omega} =
\frac{3}{16\pi} \; \left[ \frac{Q^2 + \frac{3}{2} Q_T^2}{Q^2+Q_T^2} 
+ \frac{Q^2 - \frac{1}{2} Q_T^2}{Q^2+Q_T^2} \cos^2\theta_{CS}
\right.}\nn\\[2 mm]
&& \quad \left. \mbox{} 
+\frac{Q_T Q}{Q^2+Q_T^2} K(x_1,x_2,Q_T/s)
\sin 2\theta_{CS} \cos \phi_{CS} 
+ \frac{1}{2}\frac{Q_T^2}{Q^2+Q_T^2} \sin^2 \theta_{CS}
\cos 2\phi_{CS} \right] \ ,
\label{Collins79}
\ea
where the function $K(x_1,x_2,Q_T/s)$ is given by
\beq
K(x_1,x_2,Q_T/s) = \frac{\int d\xi_1\int d\xi_2\; \delta\left(
  (\xi_1-x_1)(\xi_2-x_2) - Q_T^2/s\right) \sum_a e_a^2 q_a(\xi_1,\mu)
  \bar{q}_{a}(\xi_2,\mu)\left(x_1^2/\xi_1^2 -x_2^2/\xi_2^2\right)}{\int
  d\xi_1\int d\xi_2\; \delta\left(
  (\xi_1-x_1)(\xi_2-x_2) - Q_T^2/s\right) \sum_a e_a^2 q_a(\xi_1,\mu)
  \bar{q}_{a}(\xi_2,\mu) \left(x_1^2/\xi_1^2 + x_2^2/\xi_2^2\right)} \ .
\eeq 
As one can see, for the annihilation contribution in the CS frame all effects of the 
partonic light-cone momentum fractions and the parton densities cancel in
$dN/d\Omega$, except for the term involving $\sin 2\theta_{CS} \cos \phi_{CS}$
associated with the ratio $W_{\Delta}/(2 W_T+W_L)$. As a consequence, 
the coefficients $\lambda$ and $\nu$ defined in Eq.~(\ref{lmnWrel}) are 
also free of any dependence on the parton distributions to this 
order~\cite{Collins-79}. One reads off:
\ba
\lambda_{CS}=\frac{Q^2-\frac{1}{2}Q_T^2}{Q^2+\frac{3}{2}Q_T^2}\; , \;\;\;\;
\nu_{CS} = \frac{Q_T^2}{Q^2+\frac{3}{2}Q_T^2} \ ,
\label{LOform}
\ea
so that 
\beq
1-\lambda_{CS}-2\nu_{CS}=0 \ ,
\eeq
which is the well-known Lam-Tung (LT) relation~\cite{Lam-78,Lam-80}. It is 
equivalent to $W_L = 2 W_{\Delta \Delta}$ and $A_0=A_2$; 
its origin is the equality of 
$\widehat{W}_2$ and $\widehat{W}_4$ in Eq.~(\ref{Whats}).
 
In the Gottfried-Jackson frame one finds that all terms in $dN/d\Omega$ 
depend on the parton densities and do not have particularly simple expressions. 
Remarkably, the LT relation continues to hold in this frame, however. 
This may be seen from Eq.~(\ref{lmnMatrix}) which readily shows that
the LT relation indeed holds in the GJ frame if it holds in the CS frame, 
and vice versa. In fact, it turns out that the LT relation holds for
any definition of the lepton pair c.m.s.\ frame to this order. 

Next, we consider the $qg \to \gamma^* q$ subprocess. 
The contributions to the partonic structure functions for $q(\xi_1) g(\xi_2)$ 
are~\cite{Lam-79}:
\ba 
\widehat{W}_1 & = & - \fapi\frac{2T_R}{\hat{s}\hat{t}}
\left[(\hat{s}-Q^2)^2+(\hat{t}-Q^2)^2 \right] \ ,\nn\\
\widehat{W}_2 & = & \fapi \frac{2 T_R s Q^2}{\hat{s}\hat{t}}
\left[2\xi_1(\xi_1+ \xi_2) + \xi_2^2 \right] \ , \nn\\
\widehat{W}_3 & = &\fapi \frac{4 T_R s Q^2}{\hat{s}\hat{t}}
\left[\xi_2^2- 2 \xi_1^2\right]  \ , \nn\\
\widehat{W}_4 & = &\fapi \frac{2 T_Rs Q^2}{\hat{s}\hat{t}}
\left[2\xi_1(\xi_1 - \xi_2) + \xi_2^2 \right] \ ,
\label{WhatsC}
\ea
where $T_R=1/2$. For the contribution from $g(\xi_1) q(\xi_2)$ one has to 
interchange $\xi_1$ and $\xi_2$, $\hat{t}$ and $\hat{u}$,
and, in addition, change the sign of $\widehat{W}_3$. 
 
From the expressions 
in~(\ref{WhatsC}) one finds that for the Compton process the
ratios $\lambda, \mu, \nu$ all depend on the parton distribution
functions. As discussed in Ref.\ \cite{Lindfors-79}, there is in fact 
no frame in which any of the azimuthal distributions for this process become 
independent of the parton densities (only for very large $Q_T$ does this become
approximately the case in the CS frame). Even if there were a frame in which 
this happened, the result would have no practical relevance, since one 
would always need to add the contribution by the annihilation process 
in the numerator and the denominator of the azimuthal ratios, which would 
spoil the cancellation of the parton distributions anyway. In fact, even the cancellation 
of the parton densities in $\lambda_{CS}$ and $\nu_{CS}$ in Eq.~(\ref{LOform})
for the annihilation process is of limited use, except in a pure flavor non-singlet 
situation where the Compton process is absent. 

However, despite the fact that $\widehat{W}_2 \neq \widehat{W}_4$, the LT relation 
does hold for the Compton process as well, and therefore {\it for the complete LO
Drell-Yan cross section}~\cite{Lam-79,Lam-80}. In the next section we will verify the LT relation 
very explicitly in the small-$Q_T$ limit. However, it is valid at LO regardless of
the value of $Q_T$. It is known to be (mildly) broken by NLO 
corrections~\cite{Mirkes:1992hu,Mirkes:1994dp}. 
Experimentally the LT relation was found to be rather strongly 
violated in the $\pi^- p$ Drell-Yan process~\cite{Falciano,Guanziroli,Conway}, in
disagreement (both in magnitude and in sign) with the slight
violation predicted at NLO. This has prompted much theoretical 
work~\cite{Brandenburg-93,Brandenburg-94,Eskola-94,Boer-99,Boer:2002ju,Lu:2004hu,Boer-04,Lu:2005rq,Gamberg:2005ip}, 
offering explanations that go beyond the framework of collinear
factorization and perturbative QCD to which we restrict 
ourselves in this paper. 

As a final point in this discussion of the LO contributions, we would like
to stress that the above discussion is not specific to the Drell-Yan process,
but also applies to similar processes like SIDIS
\cite{Hagiwara,Meng:1995yn,Nadolsky-99,Nadolsky-00} and
back-to-back hadron production in two-jet events in electron-positron
annihilation \cite{Soper:1982wc}, when
appropriate definitions of the frames and coordinate axes are used.

\section{Small-$Q_T$ limit of the structure functions $W_T$, $W_L$, 
$W_\Delta$, $W_{\Delta\Delta}$ at LO}

The LO results presented in the previous section 
should provide good approximations for the angular distributions in the 
Drell-Yan process at large transverse momentum, 
$Q_T\sim Q$. We shall now investigate their behavior at small $Q_T$. 
This is extracted most conveniently by using the expansion~\cite{Meng:1995yn}
\beq
\delta\left(  (1-z_1)(1-z_2) - Q_T^2/\hat{s}\right)=
\frac{\delta(1-z_1)}{(1-z_2)_+}+\frac{\delta(1-z_2)}{(1-z_1)_+}-
\delta(1-z_1)\delta(1-z_2) \ln \rho^2\,+\,{\cal O}(\rho^2)
\label{deltaexp}
\eeq
for the delta-function in Eq.\ (\ref{Wwdef}), where $z_i\equiv x_i/\xi_i$. 
Here, the ``plus''-distributions are defined as usual for an integral 
from $x$ to 1 as
\begin{equation}
\int_x^1 \frac{dz}{z} \frac{f(z)}{(1-z)_+}=
\int_x^1 \frac{dz}{z} \frac{f(z)-f(1)}{1-z}+f(1)\ln\frac{1-x}{x} \ ,
\end{equation} 
for any suitably regular function $f$. Inserting the structure
functions $\widehat{W}_i$ of Eqs.~(\ref{Whats}) and (\ref{WhatsC}) into 
Eqs.~(\ref{fromitoCS}), and expanding for small $Q_T$ 
with the help of~(\ref{deltaexp}), one finds in the CS frame:
\ba
W_{T,CS} & = & \fapi \frac{1}{\rho^2} 
\left[ -C_F\left(2 \ln \rho^2+3\right) q(x_1)
  \bar{q}(x_2) + q(x_1) \left(P_{qq} \otimes
    \bar{q}\right)(x_2)
+ \left(P_{qq} \otimes q\right)(x_1) \bar{q}(x_2) \right. \nn\\
&&\hspace*{54.75mm}\left. + \, q(x_1) \left(P_{qg} \otimes
    g\right)(x_2)
+ \left(P_{qg} \otimes g\right)(x_1) \bar{q}(x_2)+{\cal
  O}(\rho^2) \right] \ ,\nn\\
W_{L,CS} & = & 2 W_{\Delta\Delta,CS} \  ,\nn\\
&=&\fapi \left[ -C_F\left(2 \ln \rho^2+3\right) q(x_1)
  \bar{q}(x_2) + q(x_1) \left(P_{qq} \otimes
    \bar{q}\right)(x_2)
+ \left(P_{qq} \otimes q\right)(x_1) \bar{q}(x_2)\right.  \nn\\
&&\hspace*{50.5mm} \left. + \, q(x_1) \left(P_{qg}' \otimes
    g\right)(x_2)
+ \left(P_{qg}' \otimes g\right)(x_1) \bar{q}(x_2)+{\cal
  O}(\rho^2) \right]\  ,\nn\\
W_{\Delta,CS} & = & \fapi \frac{1}{\rho} \left[ q(x_1)
  \left(\tilde{P}_{qq} \otimes
    \bar{q}\right)(x_2)
- \left(\tilde{P}_{qq}\otimes q\right)(x_1) \bar{q}(x_2) \right. \nn\\[2mm]
&&\hspace*{5.75mm}\left. + \, q(x_1) \left(\tilde{P}_{qg} \otimes
    g\right)(x_2)
-\left(\tilde{P}_{qg} \otimes g\right)(x_1) \bar{q}(x_2)+{\cal
  O}(\rho^2)  \right] \ ,
\label{CSsmallQT}
\ea
where in each case we have kept the terms with the leading-power 
behavior at small $Q_T$. Furthermore, we have for notational simplicity 
suppressed the factorization scale $\mu\sim Q$ in the parton distributions,
as well as the sums over flavors, and we have defined the usual convolutions
\beq
\left({\cal P}\otimes f\right)(x_1) \equiv \int_{x_1}^1 \frac{dx}{x}
  {\cal P}(x) f\left(\frac{x_1}{x}\right) \ ,
\eeq
with ${\cal P}$ variously one of the well-known~\cite{ap} LO splitting functions
\beq
P_{qq}(x) = C_F\left[ \frac{1+x^2}{(1-x)_+} +
\frac{3}{2} \delta(1-x) \right] \ , \quad
P_{qg}(x) = T_R \left[ x^2 + (1-x)^2\right] \ ,
\label{pqq}
\eeq
or one of
\beq
P_{qg}'(x)\equiv P_{qg}(-x) \ , \quad 
\tilde{P}_{qq}(x) \equiv C_F(1+x) \ , \quad 
\tilde{P}_{qg}(x) \equiv T_R(1-2 x^2)\ .
\label{pqqt}
\eeq
Note that one could cancel the term $-3 C_F q(x_1) \bar{q}(x_2)$ 
in $W_{T,CS}$ against the contributions by the 
$3C_F  \delta(1-x)/2$ term in the splitting function $P_{qq}$ in Eq.~(\ref{pqq}). 
We have however kept the term in order to have the full splitting function 
in~(\ref{pqq}), and also because the term $\,-C_F(2 \ln \rho^2+3)\,$  
is the well-known first-order contribution to the Sudakov form factor. 
We also note that we could have first taken the small-$Q_T$ limit
of the structure functions $W_{1,2,3,4}$ in Eqs.\ (\ref{Wwdef}),~(\ref{Whats})
and then inserted the result into~(\ref{fromitoCS}). In that case, 
since all the $W_i$ have the overall power $Q_T^{-2}$ and may be
multiplied by powers of $Q_T$ in the transformation~(\ref{fromitoCS}),
it would have been crucial to keep the first sub-leading term
proportional to $\rho^0$ in the $W_i$. Otherwise, one would obtain an incorrect
result.

As can be seen in Eqs.~(\ref{CSsmallQT}), 
the annihilation process makes a logarithmic contribution
to each of the structure functions $W_T$, $W_L$, and $W_{\Delta \Delta}$,
on top of their nominal power in $Q_T$. The Compton process, on the other
hand, does not produce this leading logarithmic behavior. It is also interesting 
to note that, unlike the other structure functions, $W_{\Delta,CS}$ does not receive 
a logarithmic contribution at all. This turns out to be a result
specific to the CS-frame. In case of $W_T$, the structure at small 
$Q_T$ is well-understood in terms of the CSS formalism~\cite{CSS85}, as 
we shall briefly review in the next section. The LO small-$Q_T$ expressions 
for the other structure functions are new. We note that the Lam-Tung relation 
$W_L = 2 W_{\Delta \Delta}$ of course still holds in~(\ref{CSsmallQT}).

The small-$Q_T$ expressions for the structure functions 
$W_{T,L,\Delta,\Delta \Delta}$ in the GJ frame can be obtained in the
same way, by using Eqs.\ (\ref{fromitoGJ}), or alternatively Eqs.\
(\ref{fromCStoGJ}), and we find:
\ba
W_{T,GJ} & = & W_{T,CS} \ , \nn\\
W_{L,GJ} & = &2 W_{\Delta \Delta,GJ}\nn \\
&=&2 \fapi \left[ -C_F\left(2 \ln \rho^2+3\right) q(x_1)
  \bar{q}(x_2) + q(x_1) \left(P_{qq}^+ \otimes
    \bar{q}\right)(x_2)
+ \left(P_{qq}^- \otimes q\right)(x_1) \bar{q}(x_2) \right.  \nn\\
&&\hspace*{52.75mm} +\left. \, q(x_1) \left(P_{qg}^{\prime\, +} \otimes
    g\right)(x_2)
+ \left(P_{qg}^{\prime\, -} \otimes g\right)(x_1) \bar{q}(x_2)+{\cal
  O}(\rho^2) \right] \ , \nn \\
W_{\Delta,GJ} & = & \fapi \frac{1}{\rho}\left[ 
-C_F\left(2 \ln \rho^2+3\right) q(x_1)
  \bar{q}(x_2) + q(x_1) \left(P_{qq}^{\; +} \otimes
    \bar{q}\right)(x_2)
+ \left(P_{qq}^{\; -} \otimes q\right)(x_1) \bar{q}(x_2) \right.  \nn\\
&&\hspace*{53.5mm} +\left. \, q(x_1) \left(\tilde{P}_{qg}^+ \otimes
    g\right)(x_2)
+ \left(\tilde{P}_{qg}^- \otimes g\right)(x_1) \bar{q}(x_2)+{\cal
  O}(\rho^2) \right] \ ,
\label{GJsmallQT}
\ea
where 
\ba
P_{qq}^{\; \pm}(x) &\equiv& P_{qq}(x)\pm \tilde{P}_{qq}(x) \ ,\nn \\
P_{qg}^{\prime\, \pm}(x) &\equiv&  P_{qg}(x)+P_{qg}'(x)\pm 2 \tilde{P}_{qg}(x)=
\left\{ 
\begin{array}{l}
4 T_R \\[2 mm]
8 T_R x^2 
\end{array}
\right. \ , \nn \\[2 mm]
\tilde{P}_{qg}^{\pm}(x) &\equiv&   P_{qg}(x)\pm\tilde{P}_{qg}(x)= 
\left\{ 
\begin{array}{l}
2 T_R (1-x) \\[2 mm]
2 T_R x (2x-1) 
\end{array}
\right.
\ , 
\ea
with $P_{qq}(x)$, $\tilde{P}_{qq}(x)$, $P_{qg}(x)$, $\tilde{P}_{qg}(x)$,
$P_{qg}'(x)$ as given above. As one can see, in the GJ frame, 
all structure functions receive logarithmic contributions at small $Q_T$.
We observe that apart from $W_{T,GJ}$ none of the functions contains 
sub-leading terms (i.e., terms non-logarithmic in $\rho$) that
involve only the usual splitting functions $P_{qq}$ and $P_{qg}$ 
of Eq.~(\ref{pqq}). Therefore, $W_L$ and $W_{\Delta \Delta}$ are not 
proportional to $W_T$ at sub-leading order for small $\rho$, not even if one
restricts to the annihilation process. The LT relation holds as before. 

Comparing the small-$Q_T$ behavior in the CS and GJ frames 
we find that the rotation between the two frames simply ``reshuffles'' the 
various ``splitting functions'' appearing in the $W_{T,L,\Delta,
\Delta\Delta}$. For example, in the annihilation contribution the
functions contributing to the sub-leading terms in $W_{L,\Delta, 
\Delta\Delta}$ are $P_{qq}$ and $\pm \tilde{P}_{qq}$
in the CS frame, but $P_{qq}\pm \tilde{P}_{qq}$ in the GJ
one. This pattern, which also extends to the Compton part,
can be verified by inspection of Eq.~(\ref{fromCStoGJ}).

We finally note that the non-standard ``splitting functions'' in the 
above expressions for the small-$Q_T$ limit are associated with the 
polarization states of the virtual photon contributing to the various 
structure functions. Only if the photons in the amplitude and its
complex conjugate both have the same, and physical (transverse), polarizations,
does one recover the ordinary DGLAP splitting functions at small $Q_T$. 
This is only the case for $W_T$, cf.\ Section~II. 

\section{Effects of resummation at small $Q_T$}

The small-$Q_T$ behavior of $W_T$ is predicted by the CSS 
formalism~\cite{CSS85} which resums its singular behavior
at $Q_T/Q\to 0$ to all orders in the strong coupling constant. 
The CSS resummation is formulated in impact-parameter space.
Schematically, keeping only those terms in the formalism that
play a role for our present study, one has
\beq
\label{crsecCSS}
W_T =  \int \frac{d^2b}{4\pi} \,
e^{i{\vec{q}_T}\cdot {\vec{b}}}\,
\sum_a e_a^2 \, q_a(x_1,b_0/b) \, \bar{q}_{a}(x_2,b_0/b) 
\, e^{S(b,Q)} \; .
\eeq
Here, $S(b,Q)$ is the Sudakov form factor, given by
\beq
S(b,Q)=-\int_{b_0^2/b^2}^{Q^2}{dk_T^2\over k_T^2}
\left[A\left(\alpha_s(k_T)\right)\ln\left({Q^2\over k_T^2}\right) 
+ B\left(\alpha_s(k_T)\right) \right] \; ,
\label{Sudakov}
\eeq
where $b_0=2 e^{-\gamma_E}$ with $\gamma_E$ the Euler constant,
and where the functions $A$ and $B$ have perturbative expansions of the form
\beq
A\left(\alpha_s\right)=
\sum_{k=1}^\infty A_k\left({\alpha_s\over\pi}\right)^k \; , \quad
B\left(\alpha_s\right)
=\sum_{k=1}^\infty B_k\left({\alpha_s\over\pi}\right)^k.
\eeq
Here we step over the fact that the $b$ integration also may require the
introduction of a nonperturbative Sudakov form factor, and simply focus on the
perturbative part. 

For our purposes we only need the coefficients $A_1$ and $B_1$, which read: 
\beq 
A_1=C_F\; , \quad   B_1 = -\frac{3}{2}C_F \; .
\eeq
The term $\propto A_1$ in the Sudakov exponent generates the leading
logarithms, which in $b$-space are of the form $\alpha_s^k \ln^{2k}(bQ)$,
corresponding to $\alpha_s^k\,\ln^{2k-1}\left(Q^2 /Q_T^2\right)/Q_T^2$
in $Q_T$-space. Next-to-leading logarithms (NLL) are generated
by the term $\propto B_1$, and by the running of the strong coupling and 
of the parton distribution functions
in~(\ref{crsecCSS}). In the CSS-resummed expression,
the factorization scale is $b_0/b$. Using the customary DGLAP evolution 
equation, the parton densities at this scale may be expressed by their values
at scale $\mu\sim Q$. Expanding 
the corresponding evolution matrix for the parton densities and
the Sudakov exponential in $\alpha_s$ and considering only the first-order 
term, we recover the expression for $W_{T,CS}$ given in Eq.~(\ref{CSsmallQT}) 
(or~(\ref{GJsmallQT})). The terms of $W_{T,CS}$ in~(\ref{CSsmallQT}) that 
involve the gluon distribution are generated by the singlet mixing in the 
evolution of the parton distributions between scales $b_0/b$ and $Q$. 
Clearly, for $W_{T,CS}$ in Eq.~(\ref{CSsmallQT}) to be reproduced by the CSS 
formalism, it is crucial that the splitting functions in its expression
are the usual DGLAP splitting functions~\cite{ap} $P_{qq}$ and $P_{qg}$ that 
are associated with the evolution of quark and anti-quark distributions.

This implies that the resummation of the structure functions 
$W_L$, $W_{\Delta}$, and $W_{\Delta\Delta}$ will not precisely follow
the CSS formalism. Taking their expressions~(\ref{CSsmallQT}) in the CS frame 
as an example, we notice that the first-order expansion of the Sudakov
form factor appears in $W_{L,CS}$ and $W_{\Delta\Delta,CS}$, indicating
that the resummed expressions for these will contain the exponential of
$S(b,Q)$, as the one for $W_T$ does, and hence have the same
leading logarithms. However, the ``splitting functions''
involved in the non-logarithmic pieces are not the usual ones, as we 
already observed in the previous section. For $W_{\Delta,CS}$, even the 
Sudakov part is absent, and the ``splitting functions''
are different again. All this means that beyond NLL the resummed
expressions for $W_{L,CS}$, $W_{\Delta,CS}$, and $W_{\Delta\Delta,CS}$ 
will not organize into the structure given in Eq.~(\ref{crsecCSS}). They
likely will have a similar structure, but contain additional terms
depending on the direction of $\bm{b}$.
To be clear, this is not a problem: while the 
presence of the true spitting functions in the leading term $W_T$ at small $Q_T$ is 
{\it required} by (collinear) factorization, 
this is not the case for the 
other structure functions. In other words, the ``splitting functions''
in these are not
associated with standard parton evolution or collinear
singularities. Therefore, the structure we find only means that the 
resummation of the $W_{L,CS}$, $W_{\Delta,CS}$, 
and $W_{\Delta\Delta,CS}$ (and likewise in the GJ frame) is more
complicated beyond the leading logarithms. We do not address their 
full NLL resummation in this work, but note that the techniques of 
Refs.\ \cite{CS81,CS82,JiMaYuan} that go beyond collinear factorization 
should prove useful for this purpose.

Even without performing the full resummation, we can make some
qualitative observations regarding the effects of resummation.
The first observation from Eqs.~(\ref{CSsmallQT}) and~(\ref{GJsmallQT}) 
is that the LT relation will not be affected by resummation.
Next, we focus on
the angular coefficients $\lambda,\mu,\nu$ or $A_0,A_1,A_2$
introduced in Eqs.~(\ref{DNdOmega}) and~(\ref{Adef}), respectively, which
are ratios of the structure functions. Because of the structure 
of the leading logarithms, we expect that in the CS frame
$\lambda$ and $\nu$, and $A_0$ and $A_2$, will be rather insensitive to 
resummation effects. Note that if the annihilation process alone contributes
(which could in principle be realized by considering a flavor non-singlet 
combination of cross sections), resummation effects cancel identically 
in $\lambda_{CS}$ and $\nu_{CS}$, and they retain their LO forms
given in Eq.~(\ref{LOform}) even after resummation. On the other hand, 
$\mu$ and $A_1$, for which there are no leading logarithms in the numerator 
because of their
absence in $W_{\Delta,CS}$, will be subject to substantial modification
by resummation. In the GJ frame all ratios 
$\lambda, \mu, \nu$ will be affected by NLL resummation effects. 
These need not be small, in particular if the overall effects of 
resummation on $W_T$ are themselves large. 

Next, we confront our findings with results of the previous 
literature. Chiappetta and Le Bellac \cite{Chiappetta-86} were the first  
to study the effects of resummation on the azimuthal asymmetries in the 
Drell-Yan process. Working in the CS frame, they argued that since the 
structure functions $W_L, W_\Delta$ and $W_{\Delta \Delta}$
are less singular than $W_T$, only the latter requires resummation. 
Therefore, they took into account resummation only in the denominators
of the $A_i$, while for the numerators they employed the
LO expressions. In this way they found large effects of 
resummation on the $A_i$. In light of our discussion above,
the neglect of resummation in the numerators of the $A_i$ is not
justified. 

Similarly, in studies~\cite{Meng:1995yn,Nadolsky-99,Nadolsky-00} of 
resummation effects in semi-inclusive DIS only the resummation of the
$\phi$-independent terms was taken into account. Specifically, resummation
was not applied in studies of the ratio $\amp{\cos 2\phi}/\amp{\cos\phi}$.
Here, a Gottfried-Jackson type of frame (one of the hadron momenta defined the 
$Z$-axis) was used. Our results above show that the numerator and denominator 
of this quantity will have a somewhat different resummation, even though
it may well turn out to be the case that the residual effects on the ratio
are small. It would be desirable to revisit this quantity in the
framework of a full NLL resummation study. This also applies to
azimuthal asymmetries in polarized scattering \cite{KoikeVogelsang}.

\section{Summary}

We conclude by summarizing our main observations. We have studied
the structure functions $W_{T,L,\Delta,\Delta\Delta}$ in the Drell-Yan 
process, on the basis of the contributions by the annihilation process
$q\bar{q}\to \gamma^* g$ and the Compton channel $qg\to \gamma^*q$. 
The structure functions $W_\Delta$ and $W_{\Delta\Delta}$ in particular
generate azimuthal asymmetries in the angular distribution of the 
produced leptons. The structure functions depend in general on the choice 
of coordinate axes, and we have investigated the results in the Collins-Soper 
and the Gottfried-Jackson frames. 

We have focused on the behavior of the structure 
functions at small transverse momentum $Q_T$ of the lepton pair. We
have recovered the known small-$Q_T$ behavior of $W_T$, in terms
of the leading-order DGLAP splitting functions and of the first-order 
expansion of the Sudakov form factor. For the other structure functions, 
which are all nominally suppressed by one or two powers of $Q_T/Q$ with 
respect to $W_T$, we have found that they, too, in general have large
logarithmic terms at small $Q_T$, whose form depends on the frame chosen. 
We are not aware that this feature was pointed out previously in the 
literature. The small-$Q_T$ structure we find for $W_{L,\Delta,\Delta\Delta}$
differs from that of $W_T$, however.
In the CS frame, $W_L$ and $W_{\Delta\Delta}$ receive large 
leading-logarithmic corrections identical to the ones in $W_T$. By 
contrast, these are absent in $W_\Delta$. In the GJ frame, all
structure functions have leading-logarithmic terms. Both frames
have in common that the subleading terms in $W_{L,\Delta,\Delta\Delta}$
are different from those in $W_T$. This implies that the next-to-leading 
logarithmic resummation at small $Q_T$ must proceed differently 
from that for $W_T$ given by the CSS formalism which uses a collinear
expansion. 

Without actually deriving the NLL resummation, we have discussed some generic 
features that we expect from it. The Lam-Tung relation, which is an exact 
property of the full leading-order contributions to the structure functions,
independent of the coordinate frame, is not affected by resummation, even
though the individual terms entering in it are affected. 
Furthermore, in the CS frame, the ratios 
$\lambda = (W_T - W_L)/(W_T+W_L)$ 
and $\nu = 2W_{\Delta\Delta}/(W_T+W_L)$ will {\em not\/} receive large 
corrections from resummation at small $Q_T$. In particular, when restricting 
to the annihilation process, resummation has no effect on the  
${\cos 2\phi}$ asymmetry. This observation is especially relevant 
for the Drell-Yan process in $\pi p$ or $p \bar{p}$ scattering.
The ratio $\mu = W_{\Delta}/(W_T+W_L)$, on the other hand, 
will be subject to considerable resummation effects, due to the lack 
of the leading-logarithmic terms in its numerator. In the GJ frame all ratios 
$\lambda, \mu, \nu$ will be affected by NLL resummation effects, which are not
necessarily small. Similar conclusions were drawn for the ratios $A_i$. 

We finally emphasize again that we hope that our study will
provide motivation for a development of the full NLL
resummation for the structure functions $W_{L,\Delta,\Delta\Delta}$.
Given the experimental information available on the Drell-Yan
process, on SIDIS, and on $e^+e^-$ annihilation, this would also
have clear phenomenological relevance. Of further interest would be
studies of the angular distributions integrated over $Q_T$~\cite{grandou}. 
Here, ``threshold-type'' logarithms may emerge~\cite{dythr1,dythr2}, whose 
structure and resummation in azimuthal distributions have so far not 
been investigated.

\begin{acknowledgments}
We thank Arnd Brandenburg, Yuji Koike, Pavel Nadolsky, Jianwei Qiu and 
Feng Yuan for useful discussions.
The research of D.B.\ has been made possible by financial
support from the Royal Netherlands Academy of Arts and Sciences.
W.V.\ is grateful to RIKEN, Brookhaven National Laboratory and the U.S.\
Department of Energy (contract number DE-AC02-98CH10886) for
providing the facilities essential for the completion of part of this work.
\end{acknowledgments}


\end{document}